%% file: main.tex
\documentclass[twocolumn]{article}
\usepackage[margin=1in]{geometry}

\usepackage{booktabs}
\usepackage{graphicx}
\usepackage{geometry}
\usepackage{tabularx}
\usepackage{bold-extra}
\usepackage{eurosym}
\usepackage{siunitx}
\usepackage{datatool}
\usepackage{biblatex}
\usepackage{hyperref}
\usepackage{ifdraft}
\usepackage{xcolor}
\usepackage{datetime}
\usepackage{mathtools}
\usepackage{texdate}
\usepackage{chngcntr}
\usepackage{xurl} 
\usepackage[title]{appendix}

\graphicspath{ {./images/} }
\DeclareSIUnit\hash{H}
\DeclareSIUnit\ton{t}
\DeclareSIUnit\year{year}
\DeclareSIUnit\TWh{\tera\watt\hour}
\DeclareSIUnit\EUR{\text{\euro}}

\newcommand{\COtwo}{CO\textsubscript{2}}
\newcommand{\COtwoe}{CO\textsubscript{2}e}
\newcommand{\mtcotwo}[1]{\SI{#1}{\mega\ton\COtwo}}
\newcommand{\ktcotwo}[1]{\SI{#1}{\kilo\ton\COtwo}}
\newcommand{\ef}[1]{\SI{#1}{\gram\COtwo\per{\kilo\watt\hour}}}
\newcommand{\efe}[1]{\SI{#1}{\gram\COtwoe\per{\kilo\watt\hour}}}
\newcommand{\hr}[1]{\SI{#1}{\mega\hps}}
\newcommand{\eff}[1]{\SI{#1}{{\mega\hps}\per{\watt}}}
\newcommand{\effinv}[1]{\SI{#1}{\joule\per{\mega\hash}}}
\newcommand{\eurkwh}[1]{\SI{#1}{\EUR\per{\kilo\watt\hour}}}
\newcommand{\hps}{\hash\per{\second}}

\newcommand{\todaterange}{(2015-07-15 to 2022-09-15)}

\title{Ethereum Emissions: A Bottom-up Estimate \\
}
\author{Kyle McDonald \\ \href{mailto:kyle@kylemcdonald.net}{kyle@kylemcdonald.net}}

\begin{document}

\initcurrdate

\interfootnotelinepenalty=10000

\DTLsetseparator{=}
\begin{filecontents*}{data.dat}
% manually updated
ethereum_market_cap=\$152B
bitcoin_market_cap=\$325B
ethereum_volume=\$5B
bitcoin_volume=\$19B

% output from GPU efficiency
efficiency_mae=\eff{0.08}

% hiveon analysis
hiveos_eff_2020_04_16=\eff{0.232}
hiveos_eff_2021_11_12=\eff{0.337}
hiveos_eff_2021_11_14=\eff{0.337}

% output from nanopool
workers_under_50=21\%
hashrate_under_50=1\%
workers_under_100=31\%
hashrate_under_100=3\%
workers_between_100_350=37\%
hashrate_between_100_350=22\%
workers_above_350=32\%
hashrate_above_350=74\%
workers_above_850=1\%
hashrate_above_850=27\%
nanopool_median_a10_hashrate=\hr{374.0}
nanopool_date=2021-10-30
nanopool_date_count=5
nanopool_median_power=\SI{647}{\watt}
nanopool_efficiency_estimate=\eff{0.38}
nanopool_identifiable_hashrate=\SI{2.39}{\tera\hps}
nanopool_identifiable_count=5992
nanopool_typical_daily_samples=69000

% output from python emissions factors
de_vries_equivalent_march=\ktcotwo{3.9}
de_vries_equivalent_october=\ktcotwo{6.0}
eff_at_end=\eff{0.48}
eff_at_start=\eff{0.17}
eff_end_date=2022-09-15
emissions_max=\eff{366}
emissions_min=\eff{237}
extra_data_regex_count=18
extra_data_region_count=13
first_efficiency=\eff{0.17}
known_by_extra_data_pct=42\%
known_by_miner_pct=48\%
krause_emissions_period_best=\mtcotwo{2.6}
krause_emissions_period_lower=\mtcotwo{1.7}
krause_emissions_period_upper=\mtcotwo{4.3}
krause_first_date=2016-01-01
krause_first_efficiency=\eff{0.15}
krause_first_efficiency_ours=\eff{0.18}
krause_last_date=2018-06-30
krause_last_efficiency=\eff{0.21}
krause_last_efficiency_ours=\eff{0.29}
last_efficiency=\eff{0.48}
marro_equivalent=\ktcotwo{14.1}
mining_pool_count=60
peak_emissions_annual_best=\mtcotwo{8}
peak_emissions_annual_lower=\mtcotwo{6}
peak_emissions_annual_upper=\mtcotwo{12}
peak_emissions_best=\ktcotwo{23}
peak_emissions_date=2022-05-13
peak_emissions_lower=\ktcotwo{16}
peak_emissions_upper=\ktcotwo{32}
peak_energy_best=\SI{29}{\TWh}
peak_energy_lower=\SI{21}{\TWh}
peak_energy_upper=\SI{40}{\TWh}
peak_power_best=\SI{3.3}{\giga\watt}
peak_power_date=2022-05-13
peak_power_lower=\SI{2.3}{\giga\watt}
peak_power_upper=\SI{4.6}{\giga\watt}
total_emissions_best=\mtcotwo{18}
total_emissions_lower=\mtcotwo{13}
total_emissions_upper=\mtcotwo{27}
total_energy_2018=\SI{9.9}{\TWh}
total_energy_2020_03_27=\SI{5.6}{\TWh}
total_energy_best=\SI{60}{\TWh}
total_energy_lower=\SI{42}{\TWh}
total_energy_upper=\SI{89}{\TWh}
unknown_blocks_pct=10\%

krause_emissions_period_upper=\mtcotwo{4.3}
emissions_min=\mtcotwo{237}
emissions_max=\mtcotwo{365}
de_vries_equivalent_march=\ktcotwo{3.9}
de_vries_equivalent_october=\ktcotwo{6.1}
marro_equivalent=\ktcotwo{14.1}
eff_at_start=\eff{0.17}
eff_at_end=\eff{0.44}
eff_end_date=2021-11-21
\end{filecontents*}
\DTLloaddb[noheader, keys={thekey,thevalue}]{mydata}{data.dat}
\newcommand{\var}[1]{\DTLfetch{mydata}{thekey}{#1}{thevalue}}

\maketitle

\begin{abstract}
The Ethereum ecosystem is maintained by a distributed global network of computers that currently require massive amounts of computational power. Previous work on estimating the energy use and emissions of the Ethereum network has relied on top-down economic analysis and rough estimates of hardware efficiency and emissions factors. In this work we provide a bottom-up analysis that works from hashrate to an energy usage estimate, and from mining locations to an emissions factor estimate, and combines these for an overall emissions estimate.

\end{abstract}

\input{body}

\input{appendix}

\printbibliography

\end{document}

%% file: body.tex
\section{Introduction}

According to the 2018 Intergovernmental Panel on Climate Change\cite{ipcc_annex_2014}, staying below 1.5{\textdegree}C warming will require ``deep emissions reductions'' in order to mitigate the worst effects of climate change. This demands an accounting of the emissions associated with all human activities, including in the world of distributed computation and finance.

The Ethereum cryptocurrency used a proof-of-work (PoW) consensus algorithm.\footnote{Ethereum phased out PoW on September 15, 2022.} PoW means that lists of new transactions (called a block) are added to the ledger (a database called the blockchain) when a computer (called the worker) guesses a large effectively-random number correctly (a process called hashing or mining). This process results in a reward for the computer operator who guesses the correct number (called the miner). PoW requires miners compete to maximize their guessing speed (hashrate) and rewards, while minimizing their expenses. If their hardware is too slow, or electricity is too expensive, mining will not be profitable.

Here we analyze the energy use and emissions of the Ethereum network using a bottom-up approach. Starting with an estimate of the total hashrate, we add an estimate of hashing efficiency, and add overhead energy usage to estimate the total energy production required. Then we estimate mining locations and emissions factors in each location to produce an emissions factor estimate for the entire network. Finally we combine the energy use and emissions estimates to produce a total emissions estimate. 

This study only estimates the historical energy use and emissions of the Ethereum network. We do not consider other environmental, social, or governance issues\cite{vries_true_2021} such as electronics waste, overall fairness of cryptocurrencies, or effects of financial deregulation. We do not compare the energy use or emissions of Ethereum to other distributed computing systems or financial networks, we do not predict future energy use or emissions, and we do not attempt to account for emissions responsibility.

All code and data for reproducing this work is available at \url{https://github.com/kylemcdonald/ethereum-energy}

\section{Previous Work}

Ethereum is the second largest cryptocurrency after Bitcoin by market capitalization---the total value of all coins is \var{ethereum_market_cap}, compared to \var{bitcoin_market_cap} for Bitcoin. Ethereum is also the third most active cryptocurrency by volume---Ethereum has a daily trade volume of \var{ethereum_volume}, with Bitcoin second at \var{bitcoin_volume}.\cite{coinmarketcap_cryptocurrency_2021} Bitcoin is also a PoW currency, and Ethereum energy estimates often draw their methodology from Bitcoin energy estimates. Energy estimates are either ``top-down'', working from the price of the currency to energy based on the economics of mining; or they are ``bottom-up'', working from hardware efficiency estimates and network hashrate measurements to total energy. Koomey\cite{koomey_estimating_2019} compares six top-down Bitcoin energy estimates, with Cambridge Bitcoin Electricity Consumption Index (CBECI)\cite{cambridge_centre_for_alternative_finance_cambridge_2021} an important recent bottom-up addition. Koomey finds that different studies take different approaches to the topic, and makes recommendations for best practices that have guided this study. Recommendations include:

\begin{itemize}
    \item Report estimates to the day
    \item Provide complete and accurate information
    \item Avoid guesses and rough estimates about underlying data
    \item Collect measured data in the field for both components and systems
    \item Build from the bottom up
    \item Properly address location variations in siting of mining facilities
    \item Explicitly and completely assess uncertainties
    \item Avoid extrapolating into the future
\end{itemize}

In addition to analyzing Bitcoin, two of the studies that Koomey reviews also analyze Ethereum: de Vries\cite{de_vries_ethereum_2021} (Digiconomist) and Krause and Tolaymat\cite{krause_quantification_2018}. Besides these Bitcoin studies, two other studies analyze Ethereum: Gallersdörfer et al.\cite{gallersdorfer_energy_2020}, and Marro and Donno\cite{marro_greennftgreen_2021}.

De Vries provides an estimated and minimum energy estimate. They use the same methodology as in their Bitcoin study, assuming that 60\% of mining revenue is spent on mining costs. De Vries also uses an emissions factor for Bitcoin to estimate Ethereum emissions. Gallersdörfer et al. makes an energy estimate but not an emissions estimate. Krause and Tolaymat provide an estimate for energy, and an upper and lower bound for emissions based on representative high and low emissions factors. Marro and Donno use a hashrate-weighted average of emissions factors across three regions, and also estimate the impact of hardware manufacturing on emissions.

These four studies (de Vries, Gallersdörfer et al., Krause and Tolaymat, Marro and Donno) are in rough agreement with each other, finding that Ethereum power during 2018 varied between \SI{1}{\giga\watt} to \SI{2.5}{\giga\watt}, around \SI{10}{\TWh} to \SI{20}{\TWh} for the year. Emissions estimates span a larger range, between \ktcotwo{5} per day and \ktcotwo{30} per day in 2018. Because these studies work top-down from miner revenue to estimating investment in electricity and hardware, they are limited by assumptions about how miners make business decisions.

In this study we try to minimize assumptions by building bottom-up with measurements and observations where possible.

\section{Energy}

Ethereum energy use can be estimated over a time period based on the equation:

\newcommand{\hashrate}{\mathit{hashrate}}
\newcommand{\hardwareoverhead}{\mathit{over}_\mathit{hw}}
\newcommand{\gridloss}{\mathit{loss}_\mathit{grid}}
\newcommand{\hashingefficiency}{\mathit{efficiency}_\mathit{hashing}}
\newcommand{\datacenteroverhead}{\mathit{over}_\mathit{dc}}
\newcommand{\powersupplyefficiency}{\mathit{efficiency}_\mathit{psu}}

\begin{equation} \label{eq:1}
\frac{\hashrate \times \hardwareoverhead \times \datacenteroverhead \times \gridloss}{\hashingefficiency \times \powersupplyefficiency}
\end{equation}

\begin{itemize}
\item $\hashrate$ is the average number of guesses per second from all workers. Typically \SI{}{megahashes\per{second}} (\hr{}) for a single worker, or \SI{}{terahashes\per{second}} (\SI{}{\tera\hps}) for the whole network.
\item $\hardwareoverhead$ is the hardware overhead for CPU, network card, and other components. This is a multiplier greater than or equal to 1.
\item $\datacenteroverhead$ is the datacenter overhead, also called power usage effectiveness (PUE). This includes cooling (fans, pumps), lighting, security, networking, but not employee transportation. This is a multiplier greater than or equal to 1.
\item $\gridloss$ is the grid loss, electricity lost during the transmission from supplier to consumer. This is a multiplier greater than or equal to 1.
\item $\hashingefficiency$ is the hashing efficiency of the hardware, as hashrate per unit of power.\footnote{Sometimes efficiency is given inverted, as energy-per-hash (for example, Joule-per-megahash or \effinv{}). We use hash-per-energy so that the value increases as efficiency increases.} We use megahashes per second per Watt (\eff{}).
\item $\powersupplyefficiency$ is the power supply efficiency for the AC-DC converter powering the hardware. This is a ratio between 0 and 1, with 1 indicating a perfectly efficient power supply.
\end{itemize}

\subsection{Hashrate}

The hashrate of the Ethereum network cannot be measured directly. It can be estimated based on the block ``difficulty'' which is the current number of guesses required to add a new block. The difficulty is automatically tweaked in proportion to how frequently new blocks are submitted. This mechanism keeps the network running at the same block rate, usually around one block every 13 seconds, even as computational power is added and removed. When more computational power is added, new blocks are submitted slightly faster, causing the difficulty to increase, stabilizing the block rate. A large difficulty corresponds to a large hashrate. Etherscan\cite{etherscanio_ethereum_2021} estimates hashrate from block difficulty, for example giving a value of \SI{643}{\tera\hps} on May 20, 2021.\footnote{Different sources give slightly different hashrate estimates. CoinWarz\cite{coinwarz_ethereum_nodate} estimates from 4-5\% lower than Etherscan before 2018, then was roughly in agreement from 2018 through mid-2019, but has estimated a consistently 5-8\% higher hashrate (and rising) since mid-2019.}

\subsection{Hashing Efficiency}

We measured hardware hashing efficiency as hashrate divided by power. Hashing efficiency is difficult to estimate without access to the hardware, because both hashrate and power are connected to a variety of factors.

\begin{figure*}[htp]
    \centering
    \includegraphics[width=\linewidth]{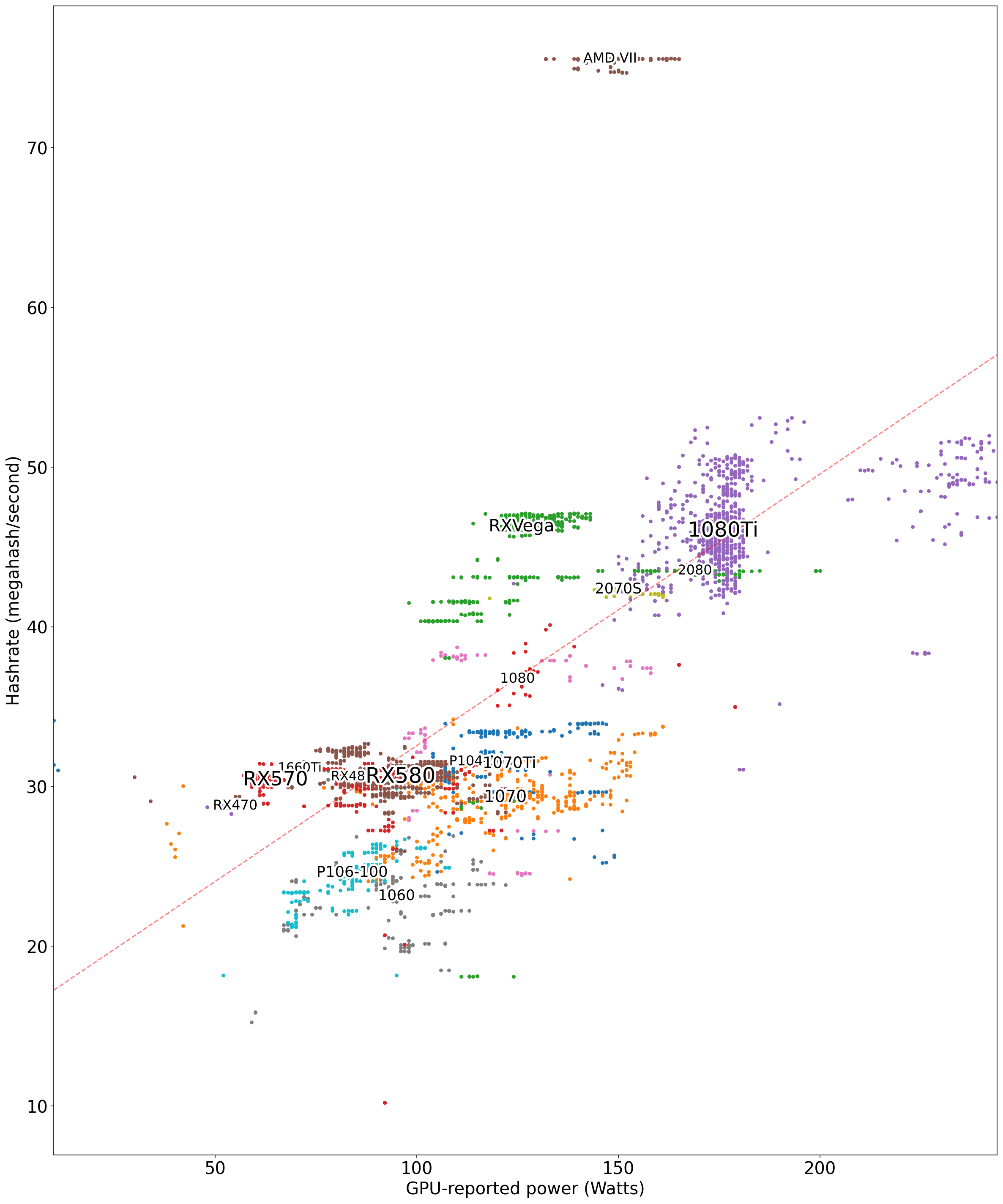}
    \caption{GPU power usage vs hashrate reported by users on MinerMonitoring, showing large variation across the same GPU models.}
    \label{fig:power-vs-hashrate}
\end{figure*}

\begin{itemize}

    \item The hardware mix used for mining Ethereum changes over time. New hardware is more efficient, and old hardware becomes either cost inefficient or can no longer hold the blockchain in memory.\cite{minerstat_dag_2021}
    
    \item The size of the blockchain may affect the hashrate. As a larger blockchain approaches hardware memory limits, the hashrate may drop. Benchmarks at one block height may not be representative of performance at another block height.\cite{walton_ethereum_2017}
    
    \item Hashrate is dependent on software configuration, with recent software sometimes hashing faster on the same hardware than older software. When faster software increases the hashrate, the power usage may not increase. Websites like MinerMonitoring\cite{minermonitoring_gpu_2021} collect benchmarks directly from users, and show broad variation in hashrate and power usage, shown in \textbf{figure \ref{fig:power-vs-hashrate}}.
    
    \item The primary hardware used for mining Ethereum are graphics processing units (GPUs, also used for gaming and machine learning). GPUs are sometimes modified (overclocked) by miners. This may increase both the hashrate and the power usage relative to out-of-the-box benchmarks.

    \item Hardware does not necessarily use its full nominal power when hashing. Hashing efficiency that is based on a hashrate benchmark and nominal power will overestimate power usage.\footnote{By analyzing benchmarks we found that power usage during hashing is around 74\% of nominal power.}
    
    \item The power self-reported by hardware may be 15\% lower\cite{mad_electron_engineering_how_2021} than the true power.

\end{itemize}

Other studies like the CBECI give a lower bound on efficiency that assumes miners are using the least efficient hardware that is still feasible and still profitable (though miners may briefly mine unprofitable coins speculatively). An upper bound on efficiency might assume that every miner is currently using the most efficient hardware available. The true efficiency is in-between.

\begin{figure*}[htp]
    \centering
    \includegraphics[width=\linewidth]{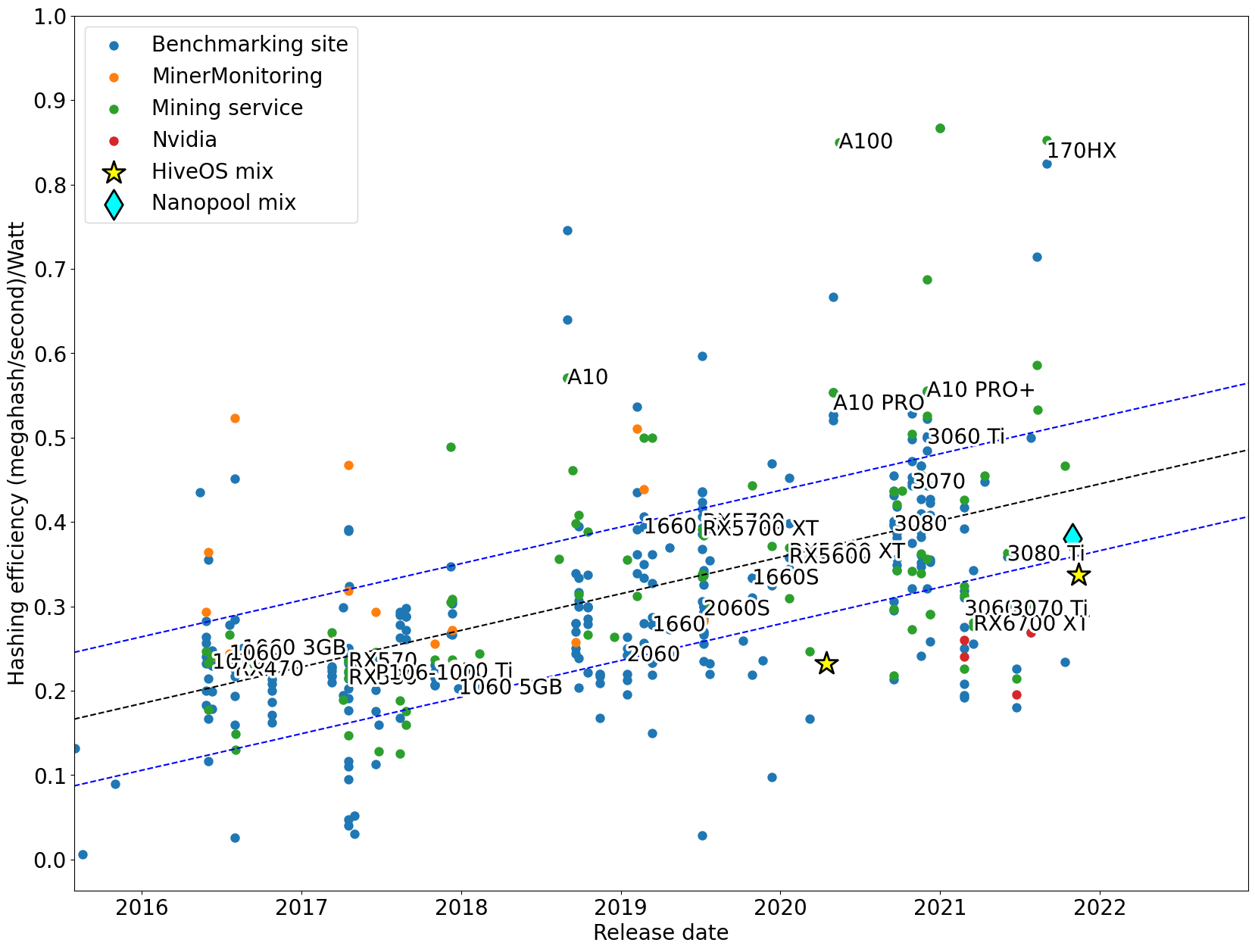}
    \caption{Plot of hardware hashing efficiency increasing over time, with hardware names drawn next to the median efficiency across multiple benchmarks.}
    \label{fig:release-vs-efficiency}
\end{figure*}

To make a best guess of hashing efficiency, we collected over 500 benchmarks from multiple sources under the assumption that mostly popular and profitable hardware is benchmarked. We compiled an additional 18 benchmarks from the median of over 9000 user-submitted benchmarks to MinerMonitoring\cite{minermonitoring_gpu_2021}. From the launch of Ethereum (2015-07-30) to the merge (\var{eff_end_date}), the typical hashing efficiency of benchmarked hardware nearly tripled from around \var{first_efficiency} to \var{last_efficiency}, as shown in \textbf{figure \ref{fig:release-vs-efficiency}}. The mean absolute difference between this trendline and the more efficient or less efficient hardware is around \var{efficiency_mae} which gives a low and high estimate.

\begin{figure*}[htp]
    \centering
    \includegraphics[width=\linewidth]{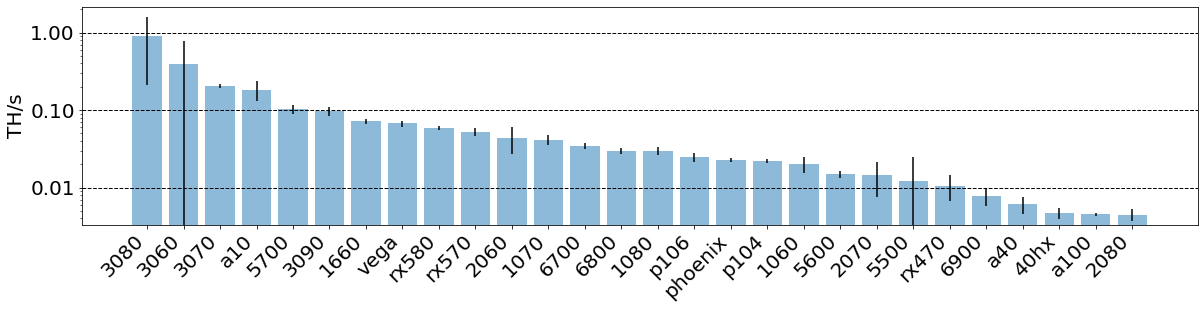}
    \caption{Total hashrate of active workers on Nanopool that match one of 28 common hardware terms, over multiple days during 2021-10 through 2021-11.}
    \label{fig:nanopool-hardware-mix}
\end{figure*}

To check our assumption that the true hardware mix is spread across a variety of devices and not just the latest hardware, we analyzed metadata from a mining pool\footnote{A small-scale Ethereum mining operation will only complete a block rarely and intermittently, so most miners join pools to take advantage of more stable revenue.}. We collected around 200,000 worker IDs (mining computer IDs) from Nanopool\cite{nanopool_nanopool_2021} per day, on \var{nanopool_date_count} different days during 2021-10 through 2021-11, with around \var{nanopool_typical_daily_samples} active workers per day (around 34\% of all active Nanopool workers).\footnote{Because workers remain registered to Nanopool even if the hardware has been retired, offline workers do not necessarily imply that there is unused hardware.} Most Nanopool worker IDs are idiosyncratic like \texttt{mpyKId18} or \texttt{Sneezy\_f559df40}. But around 8,000 IDs include hardware identifiers like \texttt{moneta\_3060Ti\_1}, or \texttt{rig5ubuntu1080*6} or even farm layout like \texttt{9throw2nda10section192x168x30x96}. Counting instances of 28 terms for common hardware, we found that a wide mix of hardware is in use, as shown in \textbf{figure \ref{fig:nanopool-hardware-mix}}.\footnote{We also found that one miner\cite{nanopool_0xa8b6_2021} with over 300 Innosilicon A10\cite{innosilicon_innosilicon_2021} units had a median performance of \var{nanopool_median_a10_hashrate}, below the advertised \hr{500}.}

To check that our most recent efficiency estimate is in the right range, we take a hashrate-weighted average of the efficiency of all identifiable Nanopool workers. Across a daily average of \var{nanopool_identifiable_count} workers representing a daily \var{nanopool_identifiable_hashrate}, we first assign each an efficiency based on the mean efficiency of hardware with the same term (for example, \texttt{3090} for the Nvidia GeForce RTX 3090 is assigned \eff{0.38}). Then we take a hashrate-weighted average, which gives us \var{nanopool_efficiency_estimate}, within our lower estimate. This efficiency sample may be skewed toward smaller operations, as larger farms are less likely to reveal their hardware mix.

As another check, we check the HiveOS\cite{hive_os_hive_2021} network statistics. HiveOS is mining monitoring software that also reports aggregate hardware statistics. For example on 2021-11-15 they report 17\% of the Nvidia models are GeForce RTX 3070 8GB, and 33\% of the AMD models are Radeon RX 580 8GB, and that 55\% of all the GPU hardware is Nvidia while the rest are AMD. We take a weighted average of the median benchmarked efficiency for all GPUs, weighted by hashrate and portion of GPU models. On 2021-11-14 we calculated an efficiency of \var{hiveos_eff_2021_11_14}, and using the the Wayback Machine\cite{hive_os_wayback_2020} we are able to calculate an efficiency for 2020-04-16 of \var{hiveos_eff_2020_04_16}. These efficiencies are below our lower estimate, but follow the same upward trend. The HiveOS statistics may represent the delay between hardware reviews and actual hardware usage, or HiveOS may not be used by the largest and highest performance mining facilities.

As a final check, we compare to a numbers from mining company Hut 8\cite{pan_ethereum_2021}. In 2021 they purchased \$30M in new GPUs (around 20-60 thousand) for a total \SI{1600}{\giga\hps} hashrate at \SI{4}{\giga\watt}. This gives \eff{0.4} measured at the plug, which implies a higher GPU-reported efficiency.

\subsection{Power Supply Efficiency}

GPU power usage benchmarks are typically reported by the GPU after the electricity has been converted from AC to DC by the power supply. Power supply units (PSUs) are inefficient. The best PSUs can be 95\% efficient while the worst are 80\% efficient or lower.

Miners aim for PSUs in the high 80\% range and many PSUs recommended to miners are rated ``80 Plus Gold`` which can perform at 90\% efficiency under a 50\% load on a 120V grid in the USA\cite{mpitziopoulos_what_2018}\cite{mpitziopoulos_best_2021}. In China and Europe this efficiency may be slightly higher on the 220V and 230V grid. The efficiency may also be lower when many cheap PSUs are bought in bulk, or pre-installed as part of a pre-built mining rig. We use power supply efficiency $\powersupplyefficiency = 0.90$

\subsection{Hardware Overhead}

While Bitcoin mining is dominated by custom computers (application-specific integrated circuits, ASICs) that are only capable of mining Bitcoin and related coins, Ethereum mining happened mainly on GPUs (with the exception of some ASICs from Innosilicon\cite{innosilicon_innosilicon_2021} and Linzhi\cite{linzhi_linzhi_2020}). Because of this, Ethereum could be mined by small-scale miners who only have a few spare GPUs that they might typically use for other applications. However, while many workers operated on one or two GPUs, most of the hashrate came from workers with closer to $6\times$ GPUs (see Appendix \ref{appendix:gpu-count}).

While hobbyist miners measure their idle power from the wall between \SI{50}{\watt} to \SI{100}{\watt}\cite{frogi5_how_2017}, a dedicated worker designed specifically for mining might use as little as \SI{20}{\watt}\cite{red_panda_mining_testing_2020}\cite{rabid_mining_ethereum_2021} for components besides the GPU.

The median GPU-reported power usage across MinerMonitoring benchmarks is \SI{109}{\watt}. Accounting for typical power supply efficiency, the total at-the-wall GPU power usage is $\SI{109}{\watt} / 0.90 = \SI{121}{\watt}$. If we assume that the typical number of GPUs per worker is 6, then total GPU-measured power usage is $6 \times \SI{121}{\watt} = \SI{726}{\watt}$ and the total worker power usage is $\SI{726}{\watt} + \SI{20}{\watt} = \SI{746}{\watt}$ so the hardware overhead is $\hardwareoverhead = \SI{746}{\watt} / \SI{726}{\watt} = 1.03$.

We can also check our per-worker GPU power usage estimate by looking at identifiable Nanopool workers. The reported hashrate divided by benchmarked efficiency equals GPU-reported power. The median power\footnote{We take the median power across all workers instead of the mean because a few workers have a very high hashrate\cite{nanopool_0xd70d_nodate}, and this may be explained\cite{majestyle1_last_2021} by miners using the same worker ID for many workers.} across all identifiable Nanopool workers is \var{nanopool_median_power}. This is close to $6 \times \SI{109}{\watt} = \SI{654}{\watt}$.

In the most efficient case where a professional miner is using 8 GPUs measured at \SI{280}{\watt} each (like the Nvidia 3090) and a 90\% efficient PSU, the GPU power at the wall is $8 \times \SI{280}{\watt} / 0.90 = \SI{2489}{\watt}$ and the additional power is \SI{20}{\watt}, $\hardwareoverhead = \SI{2509}{\watt} / \SI{2489}{\watt} = 1.01$.

In the most inefficient case where a hobbyist miner is using a single GPU measured at \SI{70}{\watt} (like the Nvidia 1660Ti), or \SI{78}{\watt} at the wall, on a computer that idles at \SI{100}{\watt}, their $\hardwareoverhead = \SI{178}{\watt} / \SI{78}{\watt} = 2.28$.

To realistically account for this inefficient case, we use data from Appendix \ref{appendix:gpu-count}, to mix the different estimates. We allocate \var{hashrate_under_100} of the hashrate to inefficient hobbyist workers, \var{hashrate_between_100_350} to the 4-6 GPU workers, and the remaining \var{hashrate_above_350} to high-efficiency workers, $\hardwareoverhead = 1.06$.

\subsection{Datacenter Overhead}

CBECI uses an estimate of 1.1 (10\% additional) based on discussions with Bitcoin miners, with a range between 1.2 (20\% additional) and 1.01 (1\% additional). Because Bitcoin is mostly mined by large-scale operations (farms), and Ethereum was mined by more workers with fewer GPUs and therefore probably by more small-scale operations, it could be argued that the datacenter overhead of these smaller operations would be higher than the overhead of a farm. Or it could alternatively be argued that a small rig has no very little marginal overhead, because they are often installed in residential contexts where overhead energy, like lighting and cooling, would be used anyway. We use a value of $\datacenteroverhead=1.1$.

\subsection{Grid Loss}

Also called grid efficiency, transmission and distribution loss, or line loss, the amount of power lost during transmission varies significantly between regions\cite{wirfs-brock_lost_2015}. Within the USA alone, grid loss varies between 2\% to 13\% across states, with an average and median around 6.5\%.

Across the regions where Ethereum is mined most often (China, USA, Northern Europe and Western Europe) grid loss varies between 5\% to 7\% and is estimated around $\gridloss = 6\%$.

\subsection{Energy Summary}

\begin{figure*}[htp]
    \centering
    \includegraphics[width=\linewidth]{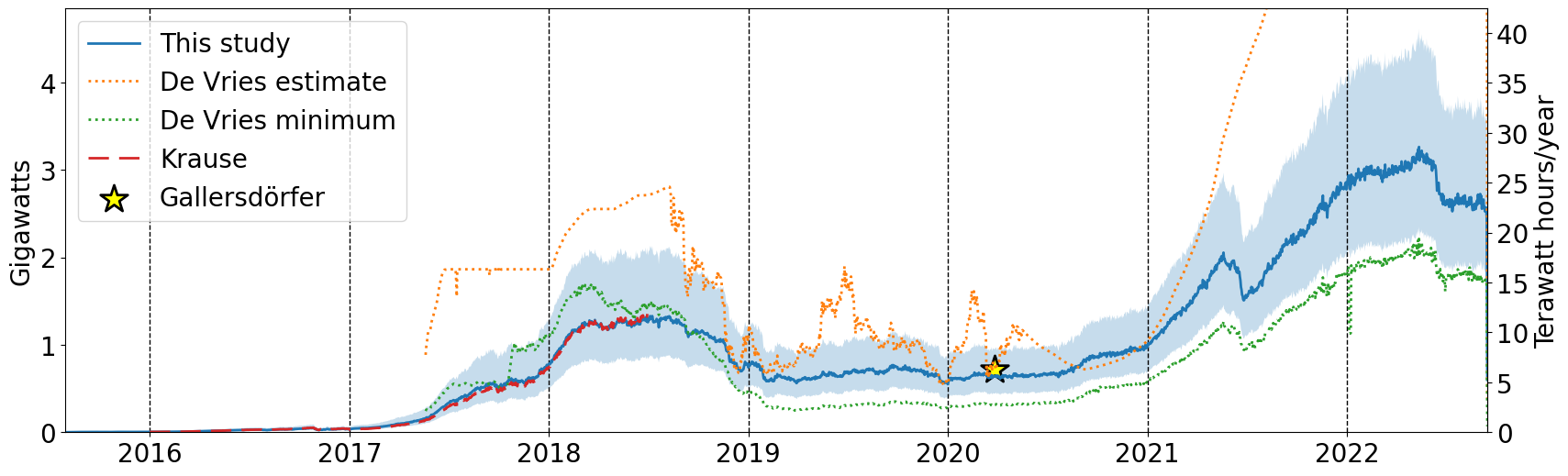}
    \caption{Ethereum network power in Gigawatts and equivalent annualized Terawatt hours per year, based on energy equation \ref{eq:1} and parameters. Shaded region shows the range between our lower and upper estimates.}
    \label{fig:energy}
\end{figure*}

In the table below, ``low'' means that the value would produce a lower energy estimate, ``high'' means the value would produce a higher energy estimate, while ``best'' is our best estimate.

\begin{center}
\begin{tabular}{ c c c c }
 & low & high & best \\
$\hashrate$ & - & - & varies \\
$\hardwareoverhead$ & 1.01 & 1.06 & 1.03 \\
$\datacenteroverhead$ & 1.01 & 1.20 & 1.10 \\
$\gridloss$ & 1.05 & 1.07 & 1.06 \\
$\hashingefficiency$ & - & - & varies \\
$\powersupplyefficiency$ & 0.95 & 0.80 & 0.90
\end{tabular}
\end{center}

With daily hashrate estimates from Etherscan\cite{etherscanio_ethereum_2021}, and our best estimate for each parameter, using equation \ref{eq:1} gives us a power estimate peaking at \var{peak_power_best} on \var{peak_power_date}, as shown in \textbf{figure \ref{fig:energy}}. Using the lower and higher parameter estimates for this day give \var{peak_power_lower} and \var{peak_power_upper}.

Applying equation \ref{eq:1} directly gives a value in units of power (Watts). Power is an instantaneous measurement. Power usage during one day might be given in Watts (average power), or as total energy during that day (power usage over time, Watt-hours or \SI{}{\watt\hour}). Other studies extrapolate to annualized energy usage, as yearly Terawatt hours or \SI{}{\TWh}.

\begin{center}
\begin{tabular}{ c c c }
 &  2020-03-27 & 2018 \\
This study & \var{total_energy_2020_03_27} & \var{total_energy_2018} \\
Gallersdörfer & \SI{6.299}{\TWh} & - \\
Krause & - & \SI{10.2}{\TWh} \\
De Vries & \SI{7.64}{\TWh} & \SI{20.8}{\TWh}
\end{tabular}
\end{center}

The results from this study mirror Krause and Tolaymat almost exactly.\footnote{Krause and Tolaymat estimate annual energy for 2018 based on data from the first half of the year, but this turned out to also be representative of the hashrate during second half of the year.} While Krause and Tolaymat do not account for datacenter overhead, hardware overhead, grid loss, and more, they estimate a lower hashing efficiency than this study: on their first date of \var{krause_first_date} they estimate \var{krause_first_efficiency} while we estimate \var{krause_first_efficiency_ours}. On their last date \var{krause_last_date} they estimate \var{krause_last_efficiency} while we estimate \var{krause_last_efficiency_ours}.

The recent increase by de Vries could be interpreted as an decrease in overall efficiency. This decrease would mostly likely be caused by a decrease in $\hashingefficiency$ or by an increase in $\hardwareoverhead$. The HiveOS data indicates that hashing efficiency has increased since 2020-04 at a similar rate to our trendline. This means that high prices may not be significantly changing the hardware mix. However, when hobbyist miners can profit from high rewards in spite of their relatively inefficient workers, their activity may decrease the overall efficiency by lowering the average $\hardwareoverhead$.

To date {\todaterange}, we estimate the Ethereum network has used between \var{total_energy_lower} and \var{total_energy_upper} with a best guess of \var{total_energy_best}.

\section{Emissions}

In order to estimate the emissions of the Ethereum network, we:

\begin{enumerate}
    \item Estimate emissions factors for popular regions where Ethereum is mined.
    \item Guess which region a block was mined in, based on block metadata.
    \item Average the emissions factors across all blocks for a given day.
    \item Multiply the daily energy usage by daily emissions factors.
\end{enumerate}

\subsection{Emissions Factors}

An emissions factor is the ratio between \COtwo{} emissions and energy production. Different kinds of electricity sources have different emissions factors. Renewables and nuclear power have very low direct emissions factors, while fossil fuels have very high emissions factors. An electricity consumer is usually connected to a mix of multiple producers, which can be treated as a weighted average of emissions factors. These energy mixes vary from place to place and over time. For example, the EPA reports that the average emissions factor in the USA in 2016 was \ef{455} and declining.\cite{us_epa_data_2020}

To analyze Ethereum, we produced a table of emissions factors for 16 electric grids (table \ref{tab:emissions-factors}). Some of these emissions factors represent a mix of multiple grids, like ``United States'' while others are within a single grid like ``Washington''.

Emissions factors were compiled from a mix of sources\footnote{Some sources that appear initially useful, like ``Carbon intensity of energy production''\cite{our_world_in_data_carbon_2019} statistics from Our World In Data, actually refer to all energy production instead of electricity alone. This includes, for example, fuel burned for transportation or heating.} including government agencies like EPA\cite{us_epa_data_2020} and EEA\cite{eea_greenhouse_2021}, corporations like BP\cite{bp_statistical_2021}, thinktanks and consultants like Shift Energy Data Portal\cite{the_shift_project_shift_2020} and Enerdata\cite{enerdata_market_2021}, as well as energy researchers like Tao et al\cite{tao_measuring_2016}. We included six years, starting in 2015 with the creation of Ethereum, through the present, 2021.

Because China has historically made up a large percentage of Ethereum mining, we take extra care calculating emissions factors there. Miners in China are known to migrate to the Southern provinces during the wet season for cheap hydropower. CBECI estimates that during the wet season in China, up to 62\% of Bitcoin mining happens in hydropower-abundant provinces like Yunnan and Sichuan. To account for this we separate out three sets of emissions factors: China as a whole (annual average), Northwest China (dry season), and South China (wet season). See Appendix \ref{appendix:emissions-factors} for details.

\subsection{\texttt{extraData} and Mining Regions}

\begin{figure*}[htp]
    \centering
    \includegraphics[width=\linewidth]{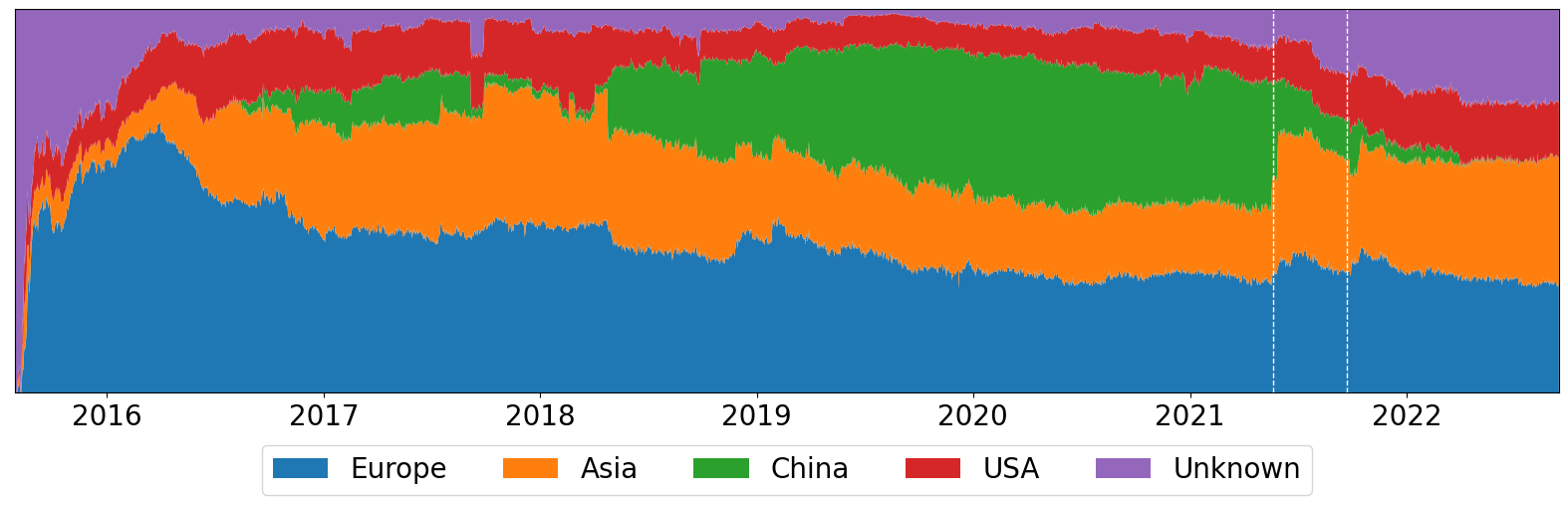}
    \caption{Distribution of mining in different regions over time based on patterns in \texttt{extraData} indicating region, and mining pool region distributions. Dashed lines for the 2021-05-21 Chinese mining ban\cite{pan_bitcoin_2021} and 2021-09-23 Chinese crypto ban\cite{gkritsi_china_2021}.}
    \label{fig:region-stackplot}
\end{figure*}

Other studies\cite{anderson_new_2016}\cite{kim_measuring_2018}\cite{etherscanio_ethereum_2021} that explore the geographic distribution of Ethereum focus on nodes instead of workers. Nodes make up the backbone of the Ethereum network, storing the blockchain and relaying new blocks to other nodes. Some workers are nodes, but if a worker is mining for a pool it is not a node. Studies based on node location may not represent mining location, and have no way of allocating hashrate. Instead, we use metadata recorded in the blockchain to estimate the regional distribution of the Ethereum hashrate.

When a new block is added to the Ethereum blockchain, it includes metadata fields, like who mined the block and when it was mined. When a worker submits a block to a mining pool server, the mining pool will add the metadata. Some mining pools use the customizable \texttt{extraData} field to report which server received the block. For example, Ethermine uses \texttt{ethermine-europe-west3} and \texttt{ethermine-europe-north1} as \texttt{extraData} values. Because miners configure their workers to connect to nearby servers (to spend more time mining and less time using the network), the \texttt{extraData} field can provide a hint about where miners are operating.

\begin{figure*}[htp]
    \centering
    \includegraphics[width=\linewidth]{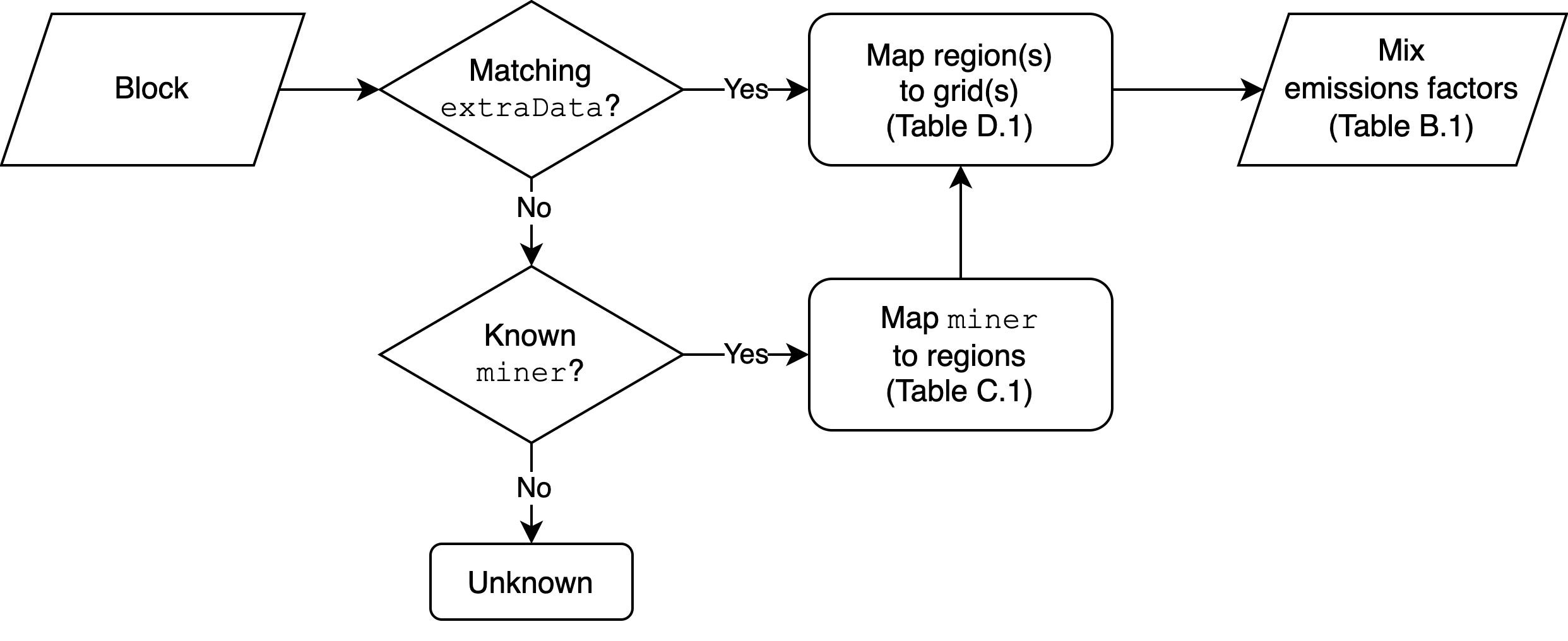}
    \caption{Process for mapping a block to an emissions factor.}
    \label{fig:mapping-flowchart}
\end{figure*}

For example, if the \texttt{extraData} matches the pattern \texttt{eu-w} then the block was likely mined in Western Europe and we assign it to the region \texttt{europe-west}. By checking \texttt{extraData} against \var{extra_data_regex_count} different patterns, we can assign a region to \var{known_by_extra_data_pct} of all blocks. Then using table \ref{tab:regions-to-grids} we treat the region as a mix of grids. For example we map the \texttt{europe-west} region to a mix of the ``Europe'', ``Netherlands'' and ``Germany'' grids. We compute the final emissions factor by taking a weighted average across the mix of grids for that year. See Appendix \ref{appendix:region-mapping} for details.

If a block cannot be mapped based on its \texttt{extraData} we try to map it based on its mining pool. For example, Huobi Mining Pool is known to operate in a mix of Europe, Asia and the USA as shown in table \ref{tab:mining-pool-regions}. So we map blocks mined by Huobi Mining Pool onto a mix of regions, \texttt{europe}, \texttt{asia} and \texttt{us}. Then we follow the same process as above to map \texttt{europe} to a mix of ``Europe'', ``Sweden'', etc., and \texttt{us} to a mix of ``United States'', ``West Virginia'' etc.

After the \var{known_by_extra_data_pct} of blocks mapped with \texttt{extraData}, an additional \var{known_by_miner_pct} of all blocks can be mapped based on \var{mining_pool_count} mining pools. This leaves only \var{unknown_blocks_pct} of blocks with a completely unknown origin. This process of mapping each block is shown in \textbf{figure \ref{fig:mapping-flowchart}}, and the resulting region distribution over time is shown in \textbf{figure \ref{fig:region-stackplot}}.

\subsection{Total Emissions}

\begin{figure*}[htp]
    \centering
    \includegraphics[width=\linewidth]{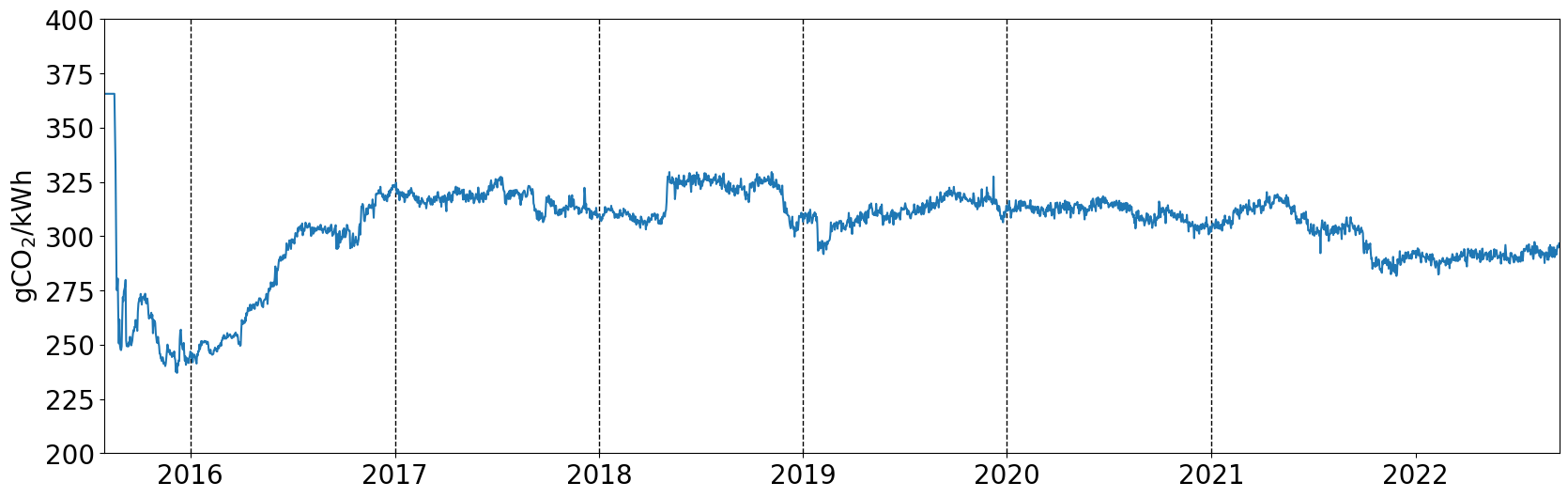}
    \caption{Ethereum electricity emissions factor \ef{}, using data from regional emissions factors weighted by region estimated from block metadata.}
    \label{fig:emissions-factors}
\end{figure*}

\begin{figure*}[htp]
    \centering
    \includegraphics[width=\linewidth]{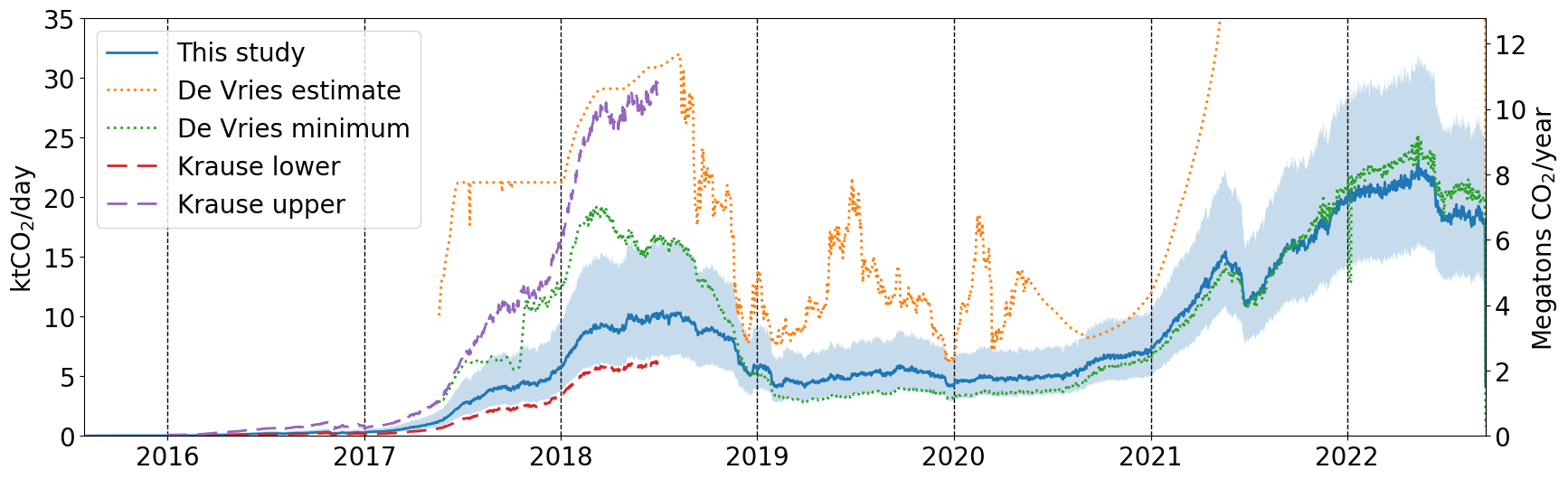}
    \caption{Ethereum emissions in \SI{}{\kilo\ton\COtwo\per{day}} and equivalent annualized \SI{}{\mega\ton\COtwo\per{year}} based on energy usage, block metadata, and regional emissions factors. Shaded region shows the range between our lower and upper estimates.}
    \label{fig:emissions}
\end{figure*}

To compute a daily emissions factor in \textbf{figure \ref{fig:emissions-factors}}, we take an average across the emissions factors of all mapped blocks for that day.\footnote{This daily emissions factor estimate is noisy for the first three weeks, so we replace that data with the median over that period.} Then to estimate total daily emissions in \textbf{figure \ref{fig:emissions}}, we multiply the emissions factor for that day by the energy used on that day. This gives us an emissions estimate peaking at \var{peak_emissions_best} on \var{peak_emissions_date}. Using the lower and higher parameter estimates for this day give \var{peak_emissions_lower} and \var{peak_emissions_upper}.

Krause and Tolaymat estimate between \mtcotwo{0.291} and \mtcotwo{1.37} total for the period from 2016-01-01 through 2018-06-30. We estimate between \var{krause_emissions_period_lower} and \var{krause_emissions_period_upper} with a best guess of \var{krause_emissions_period_best}. This discrepancy is due to an inappropriate methodology used by Krause and Tolaymat. They compute a daily ``energy-per-coin'', take the median of these values, then multiply this median by the total number of coins, and finally multiply by an emissions factor. If we simplify their methodology by computing a daily ``emissions-per-day'' then summing, their numbers give us a lower bound of \mtcotwo{1.52} for \ef{193} and an upper bound of \mtcotwo{7.21} for \ef{914}.

De Vries uses a fixed emissions factor of \ef{475} based on a location distribution of Bitcoin miners reported by Hileman and Rauchs\cite{hileman_2017_2017}. De Vries started providing annualized emissions estimates around 2021-03-08 at \mtcotwo{12.05}, with a more recent estimate on 2021-10-08 at \mtcotwo{36.41}. Our estimate is around one third on 2021-03-08 at \var{de_vries_equivalent_march}, and around one sixth at \var{de_vries_equivalent_october} on 2021-10-08. Our lower emissions estimate is partially due to our lower emissions factor estimate, but primarily due to our lower energy estimate.

Marro and Donno estimate \SI{67.80}{\kilo\ton\COtwoe} average daily emissions for the period 2021-04-29 through 2021-05-05, while we estimate \var{marro_equivalent}. Their estimate may be treated as a kind of upper bound, because it is a top-down estimate based on the assumption that miners spend all their revenue on hardware and electricity with zero profit. This discrepancy may also be related to their low estimate of hardware efficiency at \eff{0.17} while most recent hardware performs around \eff{0.3} or better.

To date {\todaterange}, we estimate the Ethereum network has emitted between \var{total_emissions_lower} and \var{total_emissions_upper} with a best guess of \var{total_emissions_best}.

\section{Future Work}

\begin{itemize}
    
    \item Some emissions are not counted here, including manufacturing emissions and greenhouse gases beyond \COtwo{}. Marro and Donno estimate that around 5\% of hardware emissions over two years can be attributed to the hardware manufacturing. Large farms sometimes require building construction and employee transportation. Mining pools also have staff and emissions.
    
    \item The hardware overhead could be more accurately represented as a fixed power usage rather than a ratio of total energy use. As more powerful GPUs are released that use more energy per card, the CPU does not also use proportionally more energy.
    
    \item Estimating the contribution of each parameter to the energy and emissions estimates would help focus future work on the highest-contribution parameters. For energy the largest source of uncertainty is $\hashingefficiency$ and hardware mix, as well as $\hardwareoverhead$. For emissions, both the emissions factors and region distribution are major sources of uncertainty.
    
    \item Mining pool region distributions have changed over time. Some pools give additional hints about miner regions by block. For example, BTC.com reports\cite{btccom_btccom_2021} over 47,000 blocks mined across 5 regions: 73\% China, 22\% EU, 5\% USA, including a breakdown between ``Shenzhen'' and ``Beijing''. Working with mining pools may provide additional location details. In particular, F2Pool jumped from providing around 8\% of the hashrate to nearly 25\% after the 2021-09-23 Chinese crypto ban\cite{gkritsi_china_2021}. This jump may come large farms migrating out of China\cite{ibc_group_ibc_2021} to the USA, Europe, South American, or into Kazakhstan. Accurately identifying their location could significantly change recent emissions factor estimates.
    
    \item Intra-year emissions factors would create a slight yearly ``wobble'' in the daily emissions estimate: peaking during China's dry season when mining electricity is powered by more fossil fuels, and dropping in the wet season when mining is powered by hydropower. This would refine and slightly lower the overall emissions for 2018, because the majority of the mining during that year happened during the wet season.
    
    \item Because electricity is sometimes cheaper in off-peak hours, leading to more mining activity, analyzing the intra-day activity trends of a pool may also reveal some information about where the users are located.
    
    \item A portion of the hashrate may rely on off-grid access to renewable energy, or on stranded energy. While there are hundreds\cite{sigalos_bitcoin_2021} of Bitcoin farms using natural gas that would otherwise be flared or vented, it is difficult to estimate the scale for Ethereum.
    
    \item If a miner operates only when there is an energy surplus from low-emissions sources, then it could be argued that they are not creating a marginal emissions increase. While this is rare, there is at least one example: miners operating in Yunnan, China around July 2018 would have to use around \SI{10}{\giga\watt} to exhaust the surplus hydropower\cite{liu_china_2021}.
    
    \item Energy terms such as grid loss $\gridloss$ or hardware overhead $\hardwareoverhead$ can vary significantly between locations and types of miners. These terms also vary over time: for example, $\hardwareoverhead$ has likely decreased with the appearance of larger farms and more powerful GPUs. As these terms vary, so does the energy use, and also the emissions. We estimate $\hardwareoverhead$ from Nanopool, but it could also be estimated from other pools that report per-miner or per-worker hashrate like SoloPool\cite{solopool_ethereum_2021}, K1Pool\cite{k1_pool_ethereum_2021}, or in archived data from Weipool\cite{weipool_weipool_2016}.
    
    \item The price of Ethereum influences the hashrate, and also the hashing efficiency. When the price increases, miners will turn on older, less efficient hardware as it becomes profitable. If the price continues to increase, some miners will use their profits to expand their operation. This implies that short-term increases in hashrate correspond to a decrease in hashing efficiency, while long-term price increases correspond to an increase in hashing efficiency.
    
    \item Geolocalized DNS activity associated with mining servers might indicate where miners are based. When a worker connects to a mining pool, it has to translate a domain name like \texttt{eth.f2pool.com} to an IP address like \texttt{172.65.210.108}. The worker will ask a DNS server for this information, and some DNS servers keep request logs. OpenDNS request statistics can be accessed using Cisco Umbrella Investigate. We checked a list of 46 mining pool servers from 11 mining pools with Umbrella Investigate in May 2021. We found that 40\% of requests originated from China, 15\% from the USA, 4\% from Brazil, 3\% from Indonesia, 3\% from Russia, plus a long tail of additional locations. Servers with \texttt{asia} in the domain name had more requests from Australia, Vietnam, Korea, and Japan. Servers with \texttt{eu} in the domain name had more requests from European countries. DNS lookups do not correspond to hashrate, and because lookups are locally cached they do not correspond worker count. Also, not all regions rely on the same mix of DNS servers.
    
    \item Hardware manufacturers like Nvidia, AMD, Linzhi, and Innosilicon may be able to provide a more precise estimate or upper bound on efficiency based on sales. Nvidia has reported \$526M\cite{leswing_nvidia_2021} in sales of crypto-specific GPUs, implying \SI{28}{\tera\hps} based on the price and hashrate reported by Hut 8\cite{pan_ethereum_2021}. A small hint regarding Linzhi is that over 500 units are identified in Nanopool worker IDs, and Linzhi writes ``For purchases of over 100 units, please contact us''.
    
    \item Our low and high estimates do not account for uncertainty in the regional distribution of mining. Emissions factors vary by more than an order of magnitude, from around \ef{40} during the west season in South China up to \ef{900} in parts of the USA. Given our network emissions factor estimates from \var{emissions_min} to \var{emissions_max}, this would imply that the total emissions could be significantly higher or lower than our high and low estimates.
    
\end{itemize}

\section{Conclusion}

We estimate Ethereum energy use based on six parameters: $\hashrate$, $\hashingefficiency$, $\hardwareoverhead$, $\datacenteroverhead$, $\gridloss$, $\powersupplyefficiency$. These parameters are either measured, estimated and checked on data, or borrowed from other studies. Our daily energy estimate generally matches studies by Gallersdörfer et al, Krause and Tolaymat, and de Vries, with the exception of a recent increase by de Vries. To date {\todaterange}, our best guess of the total energy use of Ethereum is \var{total_energy_best}.

We estimate Ethereum emissions by mapping each block to a distribution of electric grids to compute a weighted average of emissions factors. We map the blocks based on their \texttt{extraData} or their mining pool. We compute a daily emissions factor by averaging all mapped blocks. Finally, we estimate emissions by multiplying the daily energy use by the daily emissions factor. Our daily emissions estimate generally matches studies by Krause and Tolaymat, de Vries (with the exception of the same recent increase), but not Marro and Donno. To date {\todaterange}, our best guess of the total emissions of Ethereum is \var{total_emissions_best}. 

Overall, we find that most previous top-down estimates of Ethereum energy use and emissions are consistent with our bottom-up estimate.

\section{Acknowledgements}

Thanks to Jonathan Koomey, Alex Taylor and Offsetra, Memo Akten, and Michel Rauchs for discussions and guidance; to Michael Auger for coordinating exploration of DNS traffic with Cisco Umbrella Investigate; and to everyone who reviewed drafts: Dr. Keolu Fox, Emi Paternostro, Joanie Lemercier, Lia Coleman, Martial Geoffre-Rouland, Mateja Sela, Matt Aimonetti, Nao Tokui, Samantha Culp, and Tom White. Thanks to Lauren McCarthy for loving support and encouragement.

%% file: appendix.tex
\begin{appendices}

\counterwithin{figure}{section}
\counterwithin{table}{section}



\section{GPU Count}
\label{appendix:gpu-count}

\begin{figure*}[htp]
    \centering
    \includegraphics[width=\linewidth]{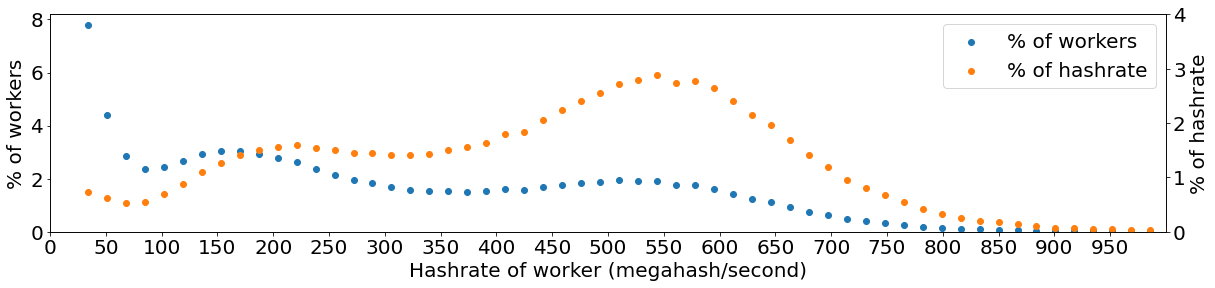}
    \caption{Analysis of the hashrate distribution of around 70,000 active Nanopool workers per day over 4 days.}
    \label{fig:nanopool-hashrate-analysis}
\end{figure*}

Analyzing the hashrate distribution of Nanopool workers reveals a pattern shown in \textbf{figure \ref{fig:nanopool-hashrate-analysis}}: around \var{workers_under_50} of workers have a hashrate under \hr{50}, and \var{workers_under_100} of workers have a hashrate under \hr{100}. This range represents the single-GPU configuration (the fastest GPUs are around \hr{100}). The next peak is around \hr{150}, likely representing the dual-GPU configuration, and the last peak is much wider around \hr{550} likely representing $6\times$ GPU configurations. While there are many few-GPU workers, workers with a hashrate under \hr{100} only represent \var{hashrate_under_100} of the total hashrate, while workers with a hashrate between \hr{100} and \hr{350} represent \var{hashrate_between_100_350} of the total hashrate, and workers over \hr{350} represent the remaining \var{hashrate_above_350}.

\section{Emissions Factors}
\label{appendix:emissions-factors}

\begin{table*}[htp]
\centering
\begin{tabular}{@{}lrrrrrrr@{}}
\toprule
Region               & 2015 & 2016 & 2017 & 2018 & 2019 & 2020 & 2021 \\ \midrule
Asia                 & 666  & 651  & 638  & 623  & 609  & 595  & 581  \\
Singapore            & 422  & 423  & 421  & 420  & 408  & 409  & 406  \\
Taiwan               & 525  & 530  & 554  & 533  & 509  & 521  & 518  \\
South Korea          & 527  & 521  & 537  & 533  & 506  & 472  & 483  \\
China                & 650  & 627  & 623  & 613  & 598  & 580  & 570  \\
China (Northwest)    & 668  & 661  & 653  & 646  & 639  & 632  & 625  \\
China (South)        & 46   & 45   & 44   & 42   & 41   & 40   & 39   \\
Europe               & 313  & 302  & 301  & 288  & 253  & 231  & 224  \\
Sweden               & 10   & 10   & 10   & 11   & 10   & 9    & 9    \\
Netherlands          & 506  & 487  & 460  & 440  & 392  & 328  & 316  \\
Germany              & 448  & 445  & 413  & 404  & 344  & 311  & 294  \\
Russia               & 326  & 327  & 325  & 325  & 323  & 301  & 307  \\
Ukraine              & 372  & 400  & 337  & 342  & 342  & 333  & 317  \\
United States        & 467  & 419  & 431  & 408  & 401  & 386  & 370  \\
United States (East) & 410  & 389  & 368  & 350  & 325  & 305  & 284  \\
United States (West) & 534  & 510  & 495  & 492  & 446  & 437  & 417  \\ \bottomrule
\end{tabular}
\caption{Regional emissions factor estimates, in \ef{}}
\label{tab:emissions-factors}
\end{table*}

An emissions factor is the ratio between \COtwo{} emissions and energy production. It is most often expressed as \ef{} or \SI{}{\kilo\gram\COtwo\per{\mega\watt\hour}} in a 0-1000 range, or as \SI{}{\ton\COtwo\per{\mega\watt\hour}} in a 0-1 range. The IPCC estimates\cite{ipcc_annex_2014} that the average emissions factor is \ef{961} for electricity generated by coal plants, and \ef{483} for gas plants, while renewables like wind and hydro emit less than \ef{10}\cite{moomaw_annex_2012} over their lifetime (due to construction and operation). An emissions factor can be used to estimate how much \COtwo{} was emitted while generating a given amount of power.

The IEA tracks and sells global emissions factors\cite{iea_emissions_2021}, but does not allow them to be published, making any derived work impossible to openly audit. To make this work auditable, we have estimated historical emissions factors for multiple regions relevant to Ethereum mining. We have also had a third party compare our emissions factors to the IEA emissions factors.

When an emissions factor is not directly available from a governmental authority, it may be estimated by:

\begin{itemize}
    \item Dividing total electricity emissions statistics by total electricity generation statistics.
    \item Taking a weighted average of the emissions factors for different energy sources in a grid (the energy mix), in proportion to the electricity generated by each source.
    \item Assuming that changes in emissions factors are relatively smooth, we may fit a line and interpolate between or extrapolate beyond known emissions factors.
\end{itemize}

Collecting and estimating emissions factors requires navigating ancient and byzantine energy agency websites, and searching for obscure presentations that include a few well-researched data points. The latest data is often a few years old. Sometimes numbers that initially look like electricity generation emissions factors turn out to actually include heating in addition to electricity, or they are projected future emissions factors accounting for power plants under construction.

We use government-reported emissions factors where possible, augmented with historical energy mix statistics, and some extrapolation and interpolation for missing data. Of all our emissions factors, 38\% are government-reported, 24\% are based on energy mix statistics, 23\% are interpolated, and 14\% are provided by non-governmental entities like the IEA.

Calculating regional emissions factors would be a straightforward process if every region had an electric grid that was completely isolated from its neighbors. In reality, grids are often connected between countries and states using a complex web of high voltage lines. These connections are managed by transmission system operators (TSOs). When electricity is purchased, it is usually bought within the scope of a TSO boundary. When the EPA gives an emissions factor of \ef{875} for the state of West Virginia in 2019, this is based on all the power plants operating in the state rather than within the TSO boundary. West Virginia is located in a RFCW TSO region, which has an emissions factor of \ef{484}.

\begin{itemize}

    \item \textbf{Asia} The IEA\cite[65]{iea_co2_2019} provides estimates for ``\ef{} of electricity'' within ``Non-OECD Asia (excluding China)'' as \ef{666} in 2015 and \ef{638} in 2017. Other years are interpolated and extrapolated. These emissions factors are designed to capture the presence of crypto mining in Southeast Asian countries like Cambodia and Vietnam\cite{redman_chinese_2018}.
    
    \item \textbf{China} We use the IEA\cite{iea_development_2020} estimates for China 2015-2020, 2021 is extrapolated. We also cross-check these numbers against the EIA\cite{us_eia_international_2021} which estimates Chinese electric generation from fossil fuels during 2015-2019 slowly dropping from 72\% to 67\%. Under the assumption that the energy mix within fossil fuels has stayed consistent, the EIA percentages match the IEA estimates within $\pm1\%$.
    
    \item \textbf{China (Northwest)} Tao et al.\cite{tao_measuring_2016} provides ``revised electric power emission factors'' for each grid region in China, approximating the complexities of emissions responsibilities implied by interprovincial power dispatching. They estimate emissions factors for all six regions from 2000-2015. We fit a line to these factors and predict 2015-2021. These predictions might be cross-referenced against Chinese Ministry of Ecology and Environment (MEE) numbers for regional grid baseline emissions factors\cite{yu_baseline_2020}. But because Tao is accounting for energy import and export, and the MEE does not, the numbers do not match. As an example, the MEE estimates Northwest China at \ef{892} during 2015-2017, while a line fit from Tao averages \ef{661} during that same period. In general, our emissions factors are low compared to the MEE estimates. Our factors are comparable to Li et al.\cite{li_life_2018} which estimates \ef{650} for 2020 while we estimate \ef{632}.
    
    \item \textbf{China (South)} We use the same methodology as \textbf{China (Northwest)} but scale our predictions based on the energy mix in the wet season. Models from Liu and Davidson\cite{liu_china_2021} indicate that in Yunnan during the wet seasons 2016-2019, coal only contributed roughly 4\% to 6\% of the total electricity generation. Given that Chinese electricity in 2017 was 70\% fossil fuels\cite{us_eia_international_2021} and mostly coal, and that the 2017 emissions factor was \ef{623}\cite[67]{iea_co2_2019} we can estimate Chinese fossil fuels to emit roughly \ef{890}, similar to the \ef{870} given by Ademe\cite{baude_key_2020} or the \ef{913} minimum from the IPCC\cite{ipcc_annex_2014}. Assuming 5\% coal in Yunnan during the wet season gives us an emissions factor of $5\%\times\ef{890}=\ef{44}$. Scaling the line fit from Tao gives us a final result of \ef{46} in 2015 dropping to \ef{39} in 2021. These factors are designed to capture the very low emissions associated with electricity generation in Sichuan and Yunnan from an abundance of hydropower. These wet season numbers are lower than the annual \ef{80} to \ef{200} range given by Li et al. for 2020 in Sichuan, and much lower than the annual \ef{225} to \ef{155} range given by a line fit from Tao et al. over 2015-2021. Both Sichuan and Yunnan are part of larger TSO regions, the Central and Southern power grids, which have much higher emissions. We use local emissions instead of TSO-wide emissions because the low price of electricity in this region indicates that it is due to an inability to efficiently export the power.
    
    \item \textbf{Europe} The EEA\cite{eea_greenhouse_2021} estimates ``greenhouse gas emission intensity of electricity generation'' for the European Union as declining from \efe{313} in 2015 to \efe{231} in 2020. While \COtwoe{} emissions cover more greenhouse gases than \COtwo{}, these numbers closely match the IEA\cite{iea_development_2020}. We extrapolate for 2021.
    
    \item \textbf{Germany}, \textbf{Netherlands} and \textbf{Sweden} We use the EEA\cite{eea_greenhouse_2021} estimates for 2015 to 2020, and extrapolate for 2021. The national borders of Sweden matches their TSO borders (Svenska kraftnät), while the Netherlands TSO TenneT extends into Germany which is served by three additional TSOs\cite{wikipedia_contributors_european_2021}.
    
    \item \textbf{Russia} BP\cite{bp_statistical_2021} provides electricity generation statistics for Russia including from oil, gas, coal, and total. According to the state-owned Russian energy corporation Gazprom\cite{gazprom_pjsc_2019}, emissions factors are \ef{400} for gas, \ef{600} for oil and \ef{845} to \ef{1020} for coal. We weight the yearly electricity generation statistics by these numbers, using the lower end \ef{845} for coal, and divide by the total generated electricity. For example, in 2018:

    \begin{align*}
    e_{gas} & = \ef{400}\times\SI{523}{\TWh} \\
    e_{oil} & = \ef{600}\times\SI{7.9}{\TWh} \\
    e_{coal} & = \ef{845}\times\SI{176}{\TWh} \\
    f_{mix} & = \frac{e_{gas}+e_{oil}+e_{coal}}{\SI{1109}{\TWh}} \\
     & = \ef{327}
    \end{align*}

    Similar to \ef{325} estimated for 2018 by Enerdata\cite{enerdata_market_2021}, reported by Carbon Footprint\cite{carbon_footprint_carbonfootprintcom_2020} from the Climate Transparency Report\cite{climate_transparency_brown_2018}. We apply this small scaling factor to the weighted estimates, and extrapolate to 2021. These numbers are comparable and lower than the Gazprom estimate \ef{358} for the earlier 2010-2016 period\cite[50]{gazprom_pjsc_2019}. This also matches the trend reported by the European Bank for Reconstruction and Development\cite{schreider_development_2011}, which predicted that Russian emissions would remain nearly unchanged between 2009-2020. Finally, percentages derived from BP's data give identical results to the EIA energy mix percentages: both give 64\% fossil fuels for 2018.
    
    \item \textbf{Singapore} We use the Energy Market Authority of the Singapore Government\cite{energy_market_authority_singapore_2021} data for 2015-2019. 2020 and 2021 are extrapolated.
    
    \item \textbf{South Korea} We use the IEA\cite{iea_development_2020} estimates for South Korea 2015-2020, 2021 is extrapolated.
    
    \item \textbf{Taiwan} We use the Taiwan Bureau of Energy 2020 Energy Statistics Handbook for 2015-2019\cite[17]{taiwan_bureau_of_energy_energy_2020} and extrapolate for 2020 and 2021.
    
    \item \textbf{Ukraine} We use the EIA fossil fuel percentage estimate from 2010-2020. We then use two equally weighted datapoints from The Climate Registry\cite{the_climate_registry_default_2021}: \ef{392} in 2010 and \ef{450} in 2011. These datapoints imply between \ef{860} to \ef{941} for fossil fuel based emissions in Ukraine. This generally matches other statistics\cite{wikipedia_contributors_energy_2021}\cite{the_shift_project_shift_2020} that report Ukraine's fossil fuel energy mix around 80\% coal and increasing since 2010. We extrapolate for 2021.
    
    \item \textbf{United States} We use the EPA eGRID data for 2018 and 2019\cite{us_epa_data_2020}. We divide the total emissions by the total generation across all regions. For example, in 2019 the US emitted \SI{1.83}{\giga\ton\COtwo} (short tons\footnote{The EPA reports \COtwo{} emissions in units of ``short tons'' equivalent to 2000 pounds, or around \SI{907}{\kilo\gram}}) for \SI{4,140}{\TWh}, around \SI{0.442}{\ton\COtwoe\per{\mega\watt\hour}} (short tons), or \ef{401}. We can use the same technique to estimate the emissions for 2016 using the eGRID 2016 Summary Tables\cite{us_epa_download_2020}. For 2020 we use EIA fossil fuel percentages. First we estimate an emissions factor for fossil fuels in 2019 as \ef{646}, given the \ef{401} overall emissions factor and 61\% fossil fuels statistic. We multiply that by the 59\% fossil fuels percentage for 2020 to get \ef{386} for 2020. We repeat this for 2015 using 2016, and for 2017 using 2016 and 2018. Finally, we extrapolate from this data to 2021.
    
    \item \textbf{United States (West)} and \textbf{United States (East)} Recent research\cite{sigalos_bitcoin_2021} indicates that 80\% Bitcoin mining in the USA happens in New York (NYUP), Kentucky (SRTV), Georgia (SRSO), Texas (ERCT) and Nebraska (MROW). We assume that Ethereum follows a similar pattern. We assign Nebraska and Texas as representing the West, and the others for the East. We use EPA eGRID data for 2018 and 2019 and eGRID 2016 Summary Tables for each of the corresponding TSO regions, and take the average across the regions. 2015, 2017, and 2020-2021 are extrapolated.

\end{itemize}

\section{Mining Regions}
\label{appendix:mining-regions}

When the region of a block cannot be estimated from its \texttt{extraData} field, we use the mining pool to estimate where it was mined. Some of the largest mining pools have multiple international servers, and others have fewer servers or have users that are geographically concentrated. We map mining pools to one or more of five regions: \texttt{asia}, \texttt{china}, \texttt{europe}, \texttt{russia}, \texttt{us}. \texttt{asia} refers to mining pools that are used broadly across Asia (including China, Japan, South Korea, and throughout Southeast Asia), while \texttt{china} is meant for pools used primarily by Chinese miners.

\begin{table}[htp]
\centering
\begin{tabular}{@{}llll@{}}
\toprule
Miner & \texttt{europe} & \texttt{us} & \texttt{asia} \\ \midrule
2Miners & 95 & 4 & 1 \\
Bw Pool & 32 & 10 & 58 \\
Coinotron & 92 & 8 &  \\
DwarfPool & 84 & 12 & 4 \\
Ethermine & 72 & 15 & 13 \\
F2Pool & 23 & 10 & 67 \\
firepool & 27 & 13 & 60 \\
Hiveon & 85 & 14 & 1 \\
Huobi Mining Pool & 49 & 25 & 26 \\
Nanopool & 94 & 6 &  \\
PandaPool & 24 & 11 & 65 \\
Spark Pool & 20 & 10 & 70 \\
UUPool & 6 & 4 & 90 \\
xnpool.cn & 32 & 14 & 54 \\
zhizhu.top & 24 & 9 & 67 \\ \bottomrule
\end{tabular}
\caption{An excerpt of mappings from mining pools to regions based on data from Silva et al\cite{silva_impact_2020}.}
\label{tab:mining-pool-regions}
\end{table}

Pool region distribution estimates are based on:

\begin{itemize}
    \item \textbf{Block propagation delay} During a one-month period from 2019-04-01 to 2021-05-02, Silva et al\cite{silva_impact_2020} logged activity on the Ethereum network using four servers. The servers were located in Portugal, Czech Republic, Taiwan, and the USA. When a block is mined by a pool in China, the block will first be reported to servers in Taiwan before it is reported to servers in the USA. For example, 94\% of Nanopool-mined blocks arrived in Portugal or Czech Republic before other regions, so we weight Nanopool as 94\% \texttt{europe}. And 90\% of UUPool-mined blocks arrived in Taiwan before other regions, so we weight UUPool as 90\% \texttt{asia}. We map 15 pools this way.
    \item \textbf{Mining Pool Stats}\cite{miningpoolstats_ethereum_2021} Regions for many current pools are listed on this tracking website. Sometimes the regions are overly general (e.g. ``global'') and can be narrowed down.
    \item \textbf{Language} Website localization options can be a clue to the user base. For example \texttt{xnpool} only offers a Chinese language interface.
    \item \textbf{Server locations} For example SpiderPool only offers servers in Hangzhou, Chengdu, and Hohhot (\texttt{eth.pool.zhizhu.top}, \texttt{eth-cd}, \texttt{eth-nmg}). While F2Pool offers servers in China, North America, and Europe (\texttt{eth.f2pool.com},  \texttt{eth-na}, \texttt{eth-eu}).
\end{itemize}

\section{Region Mapping}
\label{appendix:region-mapping}

\newcommand{\diag}[1]{\multicolumn{1}{l}{\rlap{\rotatebox{60}{#1}~}}} 

\begin{table*}[!htpb]
\centering
\begin{tabular}{@{}lllllllllllllllll@{}}
label & \diag{Asia} & \diag{Singapore} & \diag{Taiwan} & \diag{South Korea} & \diag{China} & \diag{China (Northwest)} & \diag{China (South)} & \diag{Europe} & \diag{Sweden} & \diag{Netherlands} & \diag{Germany} & \diag{Russia} & \diag{Ukraine} & \diag{United States} & \diag{United States (East)} & \diag{United States (West)} \\
\texttt{asia} & \multicolumn{1}{r}{10} & \multicolumn{1}{r}{2} & \multicolumn{1}{r}{1} & \multicolumn{1}{r}{3} & \multicolumn{1}{r}{16} & \multicolumn{1}{r}{32} & \multicolumn{1}{r}{32} &  &  &  &  &  &  &  &  &  \\
\texttt{singapore} &  & \multicolumn{1}{r}{1} &  &  &  &  &  &  &  &  &  &  &  &  &  &  \\
\texttt{taiwan} &  &  & \multicolumn{1}{r}{1} &  &  &  &  &  &  &  &  &  &  &  &  &  \\
\texttt{seoul} &  &  &  & \multicolumn{1}{r}{1} &  &  &  &  &  &  &  &  &  &  &  &  \\
\texttt{china} &  &  &  &  & \multicolumn{1}{r}{1} & \multicolumn{1}{r}{2} & \multicolumn{1}{r}{2} &  &  &  &  &  &  &  &  &  \\ \midrule
\texttt{europe} &  &  &  &  &  &  &  & \multicolumn{1}{r}{2} & \multicolumn{1}{r}{6} & \multicolumn{1}{r}{2} & \multicolumn{1}{r}{1} &  &  &  &  &  \\
\texttt{europe-west} &  &  &  &  &  &  &  & \multicolumn{1}{r}{2} &  & \multicolumn{1}{r}{6} & \multicolumn{1}{r}{1} &  &  &  &  &  \\ 
\texttt{europe-north} &  &  &  &  &  &  &  & \multicolumn{1}{r}{2} & \multicolumn{1}{r}{6} &  & \multicolumn{1}{r}{1} &  &  &  &  &  \\ \midrule
\texttt{russia} &  &  &  &  &  &  &  &  &  &  &  & \multicolumn{1}{r}{3} & \multicolumn{1}{r}{1} &  &  &  \\
\texttt{ukraine} &  &  &  &  &  &  &  &  &  &  &  & \multicolumn{1}{r}{1} & \multicolumn{1}{r}{3} &  &  &  \\ \midrule
\texttt{us} &  &  &  &  &  &  &  &  &  &  &  &  &  & \multicolumn{1}{r}{2} & \multicolumn{1}{r}{5} & \multicolumn{1}{r}{3} \\
\texttt{us-east} &  &  &  &  &  &  &  &  &  &  &  &  &  & \multicolumn{1}{r}{1} & \multicolumn{1}{r}{4} &  \\
\texttt{us-west} &  &  &  &  &  &  &  &  &  &  &  &  &  & \multicolumn{1}{r}{1} &  & \multicolumn{1}{r}{4}
\end{tabular}
\caption{Mapping from \texttt{extraData} regions and mining pool regions to electric grids.}
\label{tab:regions-to-grids}
\end{table*}

In order to convert \texttt{extraData} regions and mining pool regions to emissions factors, we need to make an informed guess. For each \texttt{extraData} region, where are the miners located that are connecting to that server? And for each mining pool region, where are the miners located that are using that mining pool?

\begin{itemize}

    \item \textbf{\texttt{asia}} From looking at the \texttt{extraData} alone, Ethereum mining in Asia is dominated by users connecting to servers in China. Around 82\% of blocks mined by an Asian server were submitted by a Chinese server. Another 10\% of blocks were submitted by a server indicating ``\texttt{asia}'' in the \texttt{extraData}. The remaining 3.5\%, 2.2\% and 1\% of blocks were mined by servers indicating \texttt{seoul}, \texttt{singapore} or \texttt{taiwan} respectively. We assume that users connecting to an \texttt{asia} server or an \texttt{asia}-based mining pool will follow this same distribution. Within China we use a $2:2:1$ ratio based on observations of Bitcoin by CBECI indicating that over the 12 month period from 2020-07-01 to 2021-07-01, mining in the Southern provinces of Sichuan and Yunnan made up 42\% of all Chinese mining activity, while mining in Xinjiang made up 36\% and other regions made up 21\%. This definition of \texttt{asia} is useful before the 2021-05-21 ban in China, but future work should update this definition to account for the new mix.

    \item \textbf{\texttt{singapore}}, \textbf{\texttt{taiwan}} and \textbf{\texttt{seoul}} We map these regions directly to their corresponding national grid.
    
    \item \textbf{\texttt{china}} We use a $2:2:1$ ratio between the Southern and Northwestern provinces and China as a whole, same as used in \texttt{\textbf{asia}}.
    
    \item \textbf{\texttt{europe}} Under the assumption that most European mining happens in places with below-average electricity prices, like Sweden (and Norway, and Iceland), we map \textbf{\texttt{europe}} evenly between these grids in a $6:1$ ratio with Europe as a whole.
    
    \item \textbf{\texttt{europe-west}} We map this primarily to Netherlands, based on statistics from cloud mining platform Mining Rig Rentals\cite{mining_rig_rentals_dagger-hashimoto_2021}. Scraping 2700 mining rigs rental listings, we found that the most common listings in Europe were located in Amsterdam (710) and Germany (563), followed by the UK (34). Because electricity in Germany is priced 50\% higher than the European average\cite{eurostat_fileelectricity_2021}, it is possible that these mining rigs are using a network proxy, and are physically located outside of Germany. CBECI has also written about apparent Bitcoin mining in Germany, ``there is little evidence of large mining operations in Germany [...] that would justify these figures. Their share is likely significantly inflated due to redirected IP addresses via the use of VPN or proxy services.''
    
    \item \textbf{\texttt{europe-north}} We map this primarily to Sweden, which has very low emissions factor and cheap electricity at 60-70\% of the European average price of \eurkwh{0.11}\cite{steitz_cryptocurrency_2018} in 2018: \eurkwh{0.065} in Sweden, \eurkwh{0.071} in Norway, and \eurkwh{0.08} in Iceland.
    
    \item \textbf{\texttt{russia}} and \textbf{\texttt{ukraine}} We assume that a portion of Russian miners connect to Ukranian servers and vice versa.
    
    \item \textbf{\texttt{us}}, \textbf{\texttt{us-east}} and \textbf{\texttt{us-west}} We balance $5:3$ in favor of \texttt{us-east} when allocating across the entire USA based on indicators from \texttt{extraData}. For the West and East we guess that 20\% is from outside the mapped regions.
    
\end{itemize}

\end{appendices}